\documentclass[manuscript,amsmath]{aastex}

\newcommand{\etal}{{\it et al. }}
\newcommand{\mas}{\mu\mathrm{as}}
\newcommand{\be}{\begin{equation}}
\newcommand{\ee}{\end{equation}}

\slugcomment{draft}
\shorttitle{Observing schedules for planet searches}
\shortauthors{Ford}

\begin{document}
\title{Choice of Observing Schedules for Astrometric Planet Searches}
\author{Eric B.\ Ford\altaffilmark{1} }

\affil{Astronomy Department,
	601 Campbell Hall, 
	University of California at Berkeley, 
	Berkeley, CA 94720-3411, USA}
\email{eford@astron.berkeley.edu}

\altaffiltext{1}{Miller Research Fellow}

\begin{abstract}
The Space Interferometry Mission (SIM) will make precise astrometric
measurements that can be used to detect planets around nearby stars.
Since observational time will be extremely valuable, it is important
to consider how the choice of the observing schedule influences the
efficiency of SIM planet searches.  We have conducted Monte Carlo
simulations of astrometric observations to understand the effects of
different scheduling algorithms.  We find that the efficiency of planet
searches is relatively insensitive to the observing schedule for most
reasonable observing schedules.  
\end{abstract}

\keywords{planetary systems -- techniques: interferometric}

\section{Introduction}

The discovery of $\sim130$ extrasolar planets via radial velocity
surveys (Butler \etal 2002 and references therein;
\url{http://exoplanets.org}) has challenged theories of planet
formation and evolution.   While radial velocity surveys are most sensitive to
planets in short-period orbits around lower main-sequence stars,
astrometric surveys are most sensitive to planets in long-period
orbits, up to the duration of the survey.  Thus, radial velocity
and astrometric surveys could be complementary for studying extrasolar
planetary systems.

The flagship astrometric survey for extrasolar planets in the next
decade is the Space Interferometry Mission, SIM
(\url{http://sim.jpl.nasa.gov/}).  Two SIM key projects will search
for low-mass extrasolar planets around nearby stars (Marcy \etal 2002;
Shao \etal 2002).  SIM is expected to make targeted observations with
a precision of $\sim1$ $\mu$as for differential astrometry.  The
relatively small number of targeted observations ($\sim 4500$ two
dimensional measurements with $1\mu$as precision over a 5 year mission
if 10\% of the time is devoted to planet searches) dictates that care
should be taken to maximize the value of the available observing time.
In this paper, we explore how the choice of observing schedule affects
the efficiency of an astrometric planet search, such as SIM.  While
the detailed results would inevitably differ for radial velocity
surveys, we expect that the main conclusions are likely to apply to
radial velocity surveys as well as other pointed astrometric surveys.
Our results will not be useful for non-pointed astrometric missions,
such as GAIA, which must adhere to a fixed scanning pattern determined
by the rotation and orbit of the satellite (Lattanzi \etal 2000).

We describe our assumptions and methods in \S 2.  In \S 3 we present
the results of our simulations.  In \S 4 we summarize our main
findings and conclusions.

\section{Methods}

The two studies of Sozzetti \etal (2002) and Ford \& Tremaine (2003)
simulated SIM observations and
arrived at similar detection criteria when stated in terms of the
scaled signal (see Eq.~2 below), despite the fact that Sozzetti
\etal (2002) assumed relative astrometry using three reference stars
while Ford \& Tremaine (2003) modeled astrometry relative to a fixed
reference frame.  Thus, we consider only absolute astrometric
measurements of the target star, neglecting potential complications
due to the parallax and proper motion of reference stars.  We expect
that our results can be simply applied to SIM's narrow angle planet
search as well as most other astrometric planet searches.

\subsection{Model Planets and Observations}

We simulated astrometric observations of many hypothetical stars.
Each star is assigned a position ($RA$, $Dec$), distance ($D$), proper
motion ($\vec{v_\perp}/D$),
mass ($M$), and a single planet.  The stellar positions are
distributed uniformly in a sphere of radius 20 parsecs centered on the
Sun, except that we reject any stars within 1 parsec of the Sun.  The
stellar velocities are drawn from a three dimensional Gaussian
distribution with mean $0$ km s$^{-1}$ and standard deviation $40$ km
s$^{-1}$ in each direction.  The stellar masses are set to $1 M_\odot$
and the stellar velocities are randomly directed in space.

Each planet is assigned a mass ($m$), orbital period ($P$), orbital
eccentricity ($e$), inclination of the orbital plane to the plane of
the sky ($i$), argument of pericenter ($\omega$), longitude of
ascending node ($\Omega$), and mean anomaly at a specified time
($M_o$).  The planetary mass and orbital period are drawn from the
mass-period distribution of Tabachnik \& Tremaine (2002), with masses
ranging from one Earth mass ($1 M_\oplus$) to ten Jupiter masses ($10
M_{Jup}$) and orbital periods ranging from $2$ days up to the duration
of the astrometric survey ($P_{SD}$).  The eccentricities 
are drawn from a
uniform distribution between $0$ and $1$.  The orbits are randomly
oriented in space.

A planet in a circular orbit will cause the star to move on the plane of the sky with a
semi-amplitude,
\be
\frac{\alpha}{''} \equiv \frac{m}{M} \frac{a}{\mathrm{AU}} \frac{\mathrm{pc}}{D}.
\ee
We present our results in terms of the ``scaled signal'',
\be
S \equiv \frac{\alpha}{\sigma_d},
\ee
where $\sigma_d$ is the single measurement precision for a one
dimensional measurement of the position of the star.  For SIM, the
single measurement precision would be the precision obtained by
combining multiple one dimensional relative delay measurements between
the target star and each reference star during a single observing
visit.  In our simulations, we set $\sigma_d = 1 \mas$, but many of
our results can be scaled to different $\sigma_d$ and our conclusions
are not sensitive to the value of $\sigma_d$.

For each star we simulated $2 N_{obs} = 96$ one dimensional
astrometric observations grouped in pairs.  Each pair of observations
occurs nearly simultaneously and is made with perpendicular baselines.
The $N_{obs}$ observation times ($t_i$) are spread over the survey
duration $P_{SD} = 10$ years (although at present the nominal SIM
lifetime in only 5 years).  We will consider several different methods
for choosing the observing times in \S 3.

\subsection{Data Analysis}

For each set of simulated observations, first, we attempt to fit a
no-planet model which includes only the star's five astrometric
parameters ($\pi$, $RA$, $Dec$, and the two components of $\vec{v}$).
The parallax, $\pi$ is the inverse of the distance to the star
(measured in parsecs).  To carry out the fit, we use the
Levenberg-Marquardt algorithm (Press \etal 1992) combined with a good
initial guess of the astrometric parameters.  By ``good'' we mean that
the initial guess for the star's distance and velocity to within
$\sim1\%$ of their true values and the star's position is the weighted
average of the observations.  

After identifying the best-fit astrometric parameters, we calculate
the usual sum of squares of the residuals ($\chi_0^2$) to evaluate the
appropriateness of the no-planet model.  If a $\chi^2$ test can reject
the no-planet model with 99.9\% confidence, then we proceed to fit a
model which includes a planet.  We attempt to find the best-fit
one-planet model using the Levenberg-Marquart algorithm with good
initial guesses ($\sim 1\%$ dispersion about the true values) for the
astrometric parameters and the planet's orbital parameters.  We hold
the star's mass fixed at its actual value.  If the $\chi^2$ statistic
for the best-fit one-planet model ($\chi_1^2$) is significantly less
than $\chi_0^2$ (the best-fit for the no-planet model) according to an
$F$-test, then we consider the planet to be detected.  The fitting is
done most efficiently by using the Thiele-Innes coordinates,
\begin{eqnarray}
X_1 & = & a_{\star} \left[ \cos \omega \sin \Omega + \sin \omega \cos \Omega \cos i \right] / D \\
Y_1 & = & a_{\star} \left[ \cos \omega \cos \Omega - \sin \omega \sin \Omega \cos i \right] / D \\
X_2 & = & a_{\star} \left[ -\sin \omega \sin \Omega + \cos \omega \cos \Omega \cos i \right] / D \\
Y_2 & = & a_{\star} \left[ -\sin \omega \cos \Omega - \cos \omega \sin \Omega \cos i \right] / D, 
\end{eqnarray}
where $a_\star$ is the star's semi-major axis.  After finding the
best-fit Thiele-Innes coordinates, we convert these to conventional
orbital elements.  The planet's semi-major axis is determined by $a =
a_\star M / m$, and the planet's mass is determined by
\begin{equation}
\frac{m/M}{\left(1+m/M\right)^{2/3} } = \frac{a_\star}{\left(\frac{G}{4\pi^2} M \right)^{1/3} P^{2/3} },
\end{equation}
where $G$ is the gravitational constant and the equation is solved
itteratively.  Since we use very good initial guesses as inputs to the
local minimization algorithm, the accuracy of our mass and orbit
determinations is optimistic.  For actual systems, it will be
necessary to perform a global search which may result in even larger
errors when estimating a planet's mass and orbital parameters.  We
have not performed a global search, as that would make it impractical
to conduct such a large number of simulations as performed in this
paper.  The ability of such global searches to converge on the correct
solution is poorly understood and worthy of separate investigation.

We repeat these calculations for hundreds of thousands of stars to
determine the fraction of planets which can be detected with the
various observing schedules considered.  We also investigate the
fraction of planets whose mass and/or orbital elements are accurately
measured.

\subsection{Observing Schedules}

The previous study of Sozzetti \etal (2002) calculated the sensitivity
of SIM for detecting extrasolar planets and included a comparison of a
few possible observing schedules.  In this study, we examine a much
larger list of possible observing schedules.

We consider several possible observing schedules:
\begin{enumerate}
\item Regular Periodic: constant spacing, $\Delta t$,
\item Golomb Ruler:  times proportional to marks on shortest known Golomb ruler with $N_{obs}$ marks,
\item Regular Power Law: times proportional to the observation number raised to a power ($t_i \sim i^{\beta}$),
%
%
\item Regular Logarithmic: times proportional to logarithm of observation number ($t_i \sim \log i$) with minimum spacing, $\Delta t_{min}$,
\item Regular Geometric: times proportional to a constant raised to the power of the observation number ($t_i \sim \beta^{i}$) with minimum spacing $\Delta t_{min}$,
\item Random Uniform: random times uniform in $t$, 
\item Random Power Law: random times uniform in $t^{\beta}$,
\item Random Logarithmic: random times uniform in $\log t$ with minimum spacing $\Delta t_{min}$, and
\item Periodic with Perturbation: constant spacing $\Delta t_0$, but with a random Gaussian perturbation with zero mean and standard deviation $\epsilon \Delta t_0$.
\end{enumerate}

A Golomb ruler is a sequence of integers which can be
considered as the distance to the next mark on a ruler such that the
distance between each pair of marks is a unique integer.  That is 
\begin{equation}
\Delta_{pq} = G_q-G_p
\end{equation}
is unique for each pair of $p$ and $q$, where $1 \le p < q \le
N_{obs}$.  The shortest known Golomb ruler for a given number of marks
can provide a good basis for designing a linear interferometer
(Robinson \& Bernstein 1967), because it includes as many distinct baselines as
possible.  We consider observing schedules based on Golomb
rulers to see if such schedules are advantageous for reconstructing
planetary orbits.  Observing schedule \#2 places observations at times $t_i = G_i
P_{SD}/N_{G}$, where $P_{SD}$ is the survey duration, 
$G_i$ is the distance to the $i$-th mark on the shortest known Golomb
ruler with $N_{obs}$ marks, and $N_G$ is the length of the Golomb
ruler.  

Observing schedules \#3-5 and \#7-9 have a single free parameter.  We
consider multiple values for these parameters.

\section{Results}

For each observing strategy we have simulated observations of many
stars with a planet to determine the efficiency of the observing
schedule for detecting planets and for measuring their masses and
orbital parameters.

\subsection{Overall Rates}

First, we compare the total number of planet detections when using
each of several different observing schedules.  In Table
\ref{TableSchedStats}, columns 2 and 3, we list the fraction of
planets which are detected for each observing schedule, averaging over
all planet masses ($1 M_\oplus-10 M_{Jup}$), orbital periods ($2$
d-$10$ yr), and other parameters.  Table \ref{TableSchedStatsTerr} is
similar to Table \ref{TableSchedStats}, but only includes planets with
masses between $1 M_\oplus$ and $20 M_\oplus$.  Despite the wide
variety of observing schedules considered, all of the schedules that
we consider detect planets at very similar rates.

Next, we consider the fraction of planets for which masses and orbits
are measured with 30\% and 10\% accuracy (Tables \ref{TableSchedStats}
and \ref{TableSchedStatsTerr}, columns 4-7).  Again, most of the
observing strategies measure masses and orbits at very similar rates.
Since the mass is a function of only the orbital period and amplitude,
measuring the mass is somewhat easier than measuring all of the
orbital parameters to the same precision.

In Figs.\ 1-5, we show the rates of detecting planets and measuring
their masses and orbits.  The upper left hand panel is for detections,
the upper right hand panel is for measuring the mass with 30\%
accuracy, the lower left panel is for measuring the orbit with 30\%
accuracy, and the lower right panel is for measuring the orbit with
10\% accuracy.  In each of these figures, the different line styles
are for simulations using different observing schedules (solid,
periodic; dotted, periodic with perturbations ($\epsilon = 0.2$);
dot-long dash, logarithmic ($\Delta t_{min}=30$ d); short dash, power
law ($\beta = 0.5$); dot-short dash, geometric ($\Delta t_{min}=30$ d);
short dash-long dash, Golomb ruler).  The various curves are often
difficult to distinguish, reflecting the fact that most of the
observing schedules that we consider have very similar efficiencies
for detecting and characterizing planets.  We now discuss each figure
in turn.

\subsection{Rate versus Scaled Signal}

\begin{figure}
\plotone{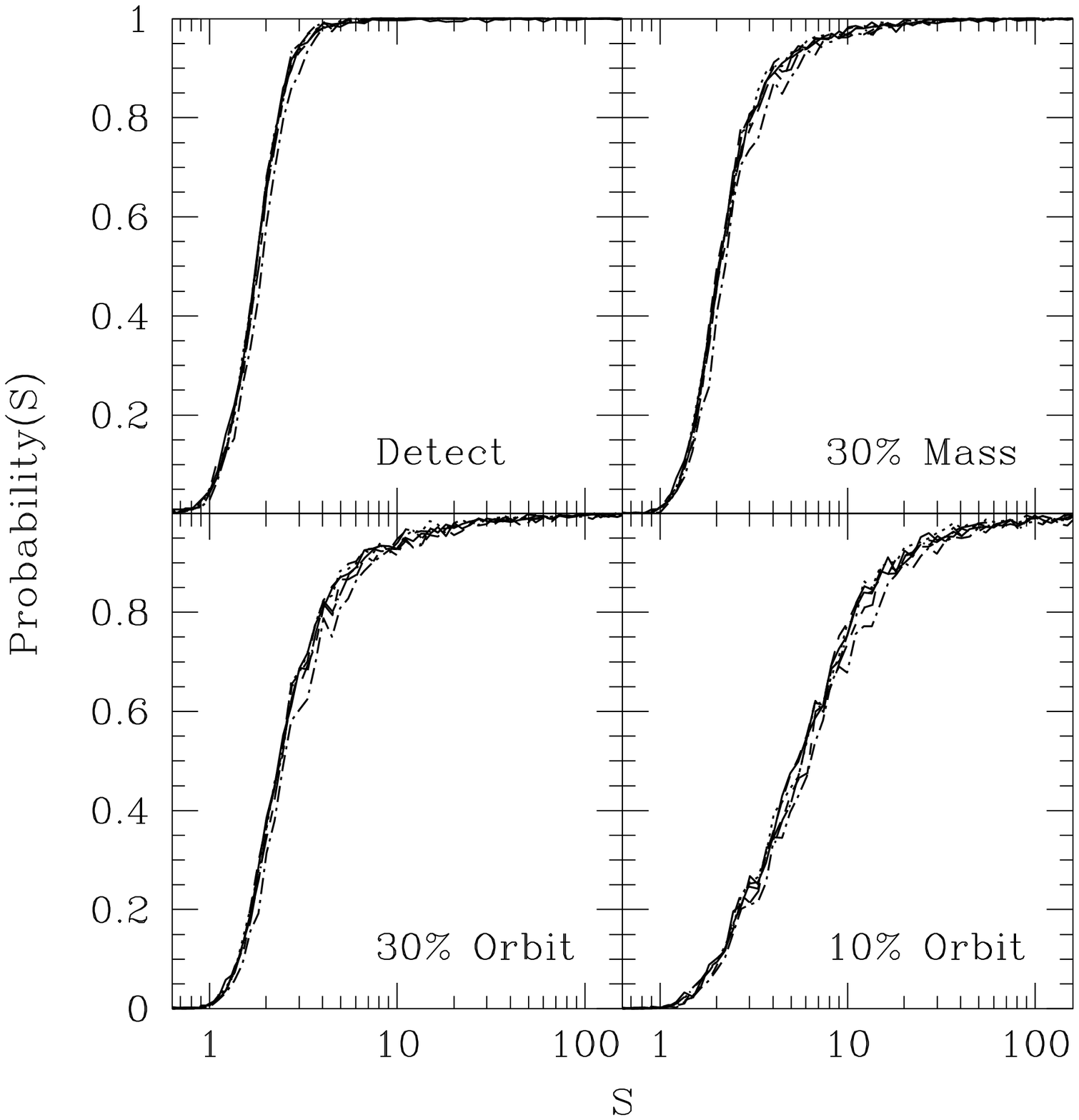}
\caption[fig1.ps]{
In each panel we show the probability for detecting a planet or
measuring its mass or orbit as a function of the scaled signal, $S$, averaging
over orbital periods, eccentricities, and other parameters.  The upper
left panel is for detecting a planet, the upper right panel is for
measuring the mass with 30\% accuracy, the lower left panel is for
measuring the orbit with 30\% accuracy, and the bottom right panel is
for measuring the orbital parameters with 10\% accuracy.  The
different line styles are for simulations using different observing
schedules (solid, periodic; dotted, periodic with perturbations
($\epsilon = 0.2$); dot-long dash, logarithmic ($\Delta t_{min}=30$ d);
short dash, power law ($\beta = 0.5$); dot-short dash, geometric
($\Delta t_{min}=30$ d); short dash-long dash, Golomb ruler).  It is
clear that our choice of observing
schedule has relatively little effect on the rates of planet detection
and characterization.
\label{ProbVsLogSForBest}}
\end{figure}

In Fig.\ \ref{ProbVsLogSForBest}, we investigate how the rates of
detecting planets and measuring their masses and orbits depend on the
scaled signal, $S$, averaging over orbital periods, eccentricities,
and other parameters.  Clearly, planets will not be detected for
sufficiently small $S$, and will be easy to detect and characterize
for sufficiently large $S$.  While there is a 50\% probability of
measuring an orbit with 10\% accuracy for a planet with a modest
scaled signal ($S\simeq6$), a significantly larger scaled signal
($S\simeq30$) is required for there to be a 95\% probability of
measuring the orbit with the same accuracy, as noted in Ford and Tremaine
(2003).  While this paper will focus on the difference between the
rates of detection and characterization for different observing
schedules, it is important to realize that all of the observing
schedules that we consider perform very similarly (see Fig.\
\ref{ProbVsLogSForBest}).

\subsection{Rate versus Orbital Period}

\begin{figure}
\plotone{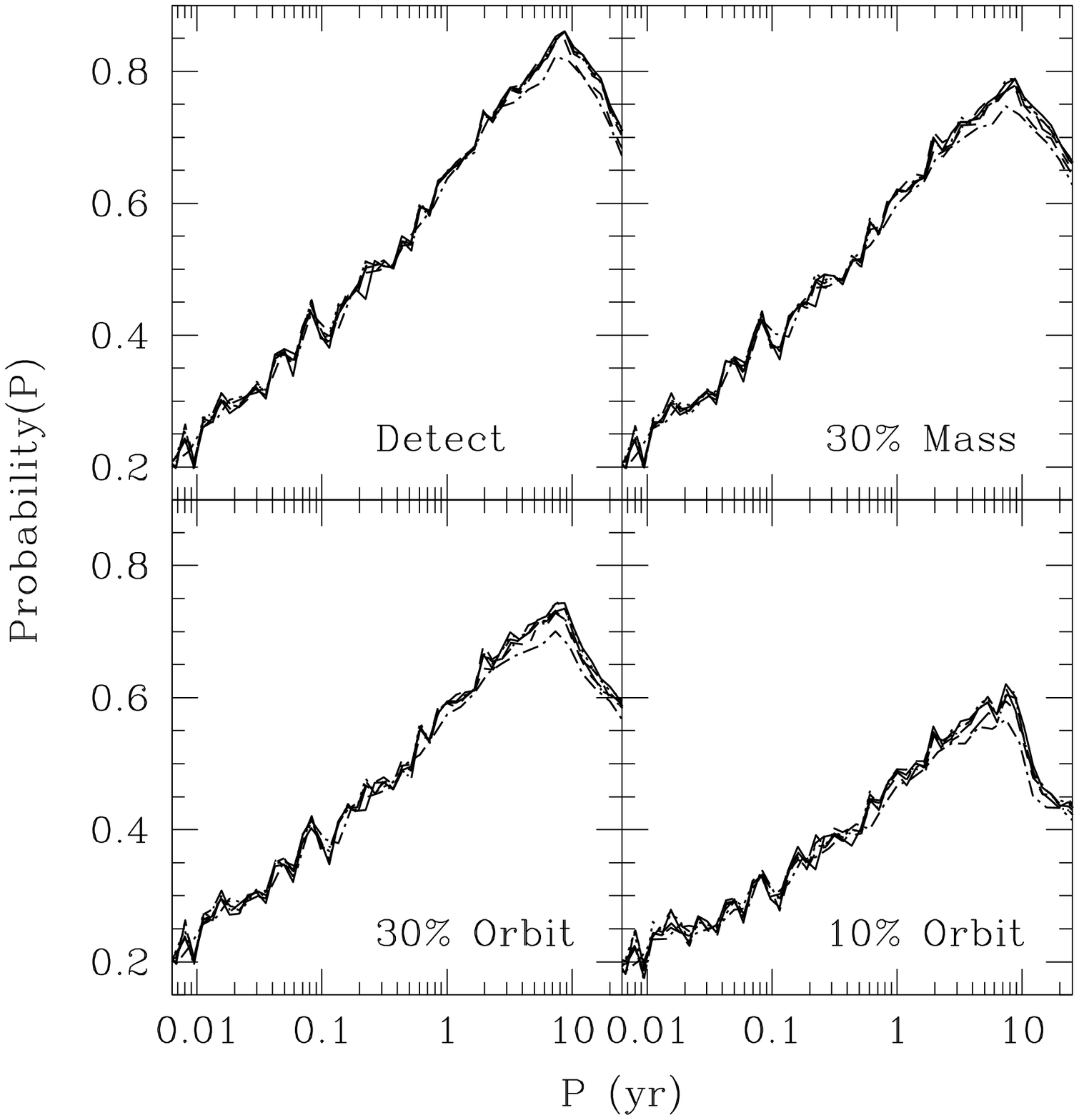}
\caption[fig2.ps]{
In each panel we show the probability for detecting a planet or
measuring its mass or orbit as a function of the orbital period, $P$,
averaging over planet mass, eccentricities, and other parameters.  The
upper left panel is for detecting a planet, the upper right panel is
for measuring the mass with 30\% accuracy, the lower left panel is for
measuring the orbit with 30\% accuracy, and the bottom right panel is
for measuring the orbital parameters with 10\% accuracy.  The
different line styles are for simulations using different observing
schedules (solid, periodic; dotted, periodic with perturbations
($\epsilon = 0.2$); dot-long dash, logarithmic ($\Delta t_{min}=30$
d); short dash, power law ($\beta = 0.5$); dot-short dash, geometric
($\Delta t_{min}=30$ d); short dash-long dash, Golomb ruler).  For all
sensible observing schedules, astrometric surveys are more sensitive
to planets with longer orbital periods, until the orbital period
approaches the duration of the astrometric survey.  Most of the
fluctuations are due to noise.
\label{ProbVsLogPForBest}}
\end{figure}

Next, we investigate how the rates of detecting planets and measuring
their masses and orbits depends on the orbital period, $P$, averaging
over a distribution of planet masses based on the observed radial
velocity planets (Tabachnik and Tremaine 2002).  Planets with very
short period orbits are typically difficult to detect, primarily due
to the smaller scaled signal, $S$ (see Fig.\ \ref{ProbVsLogPForBest}).

\begin{figure}
\plotone{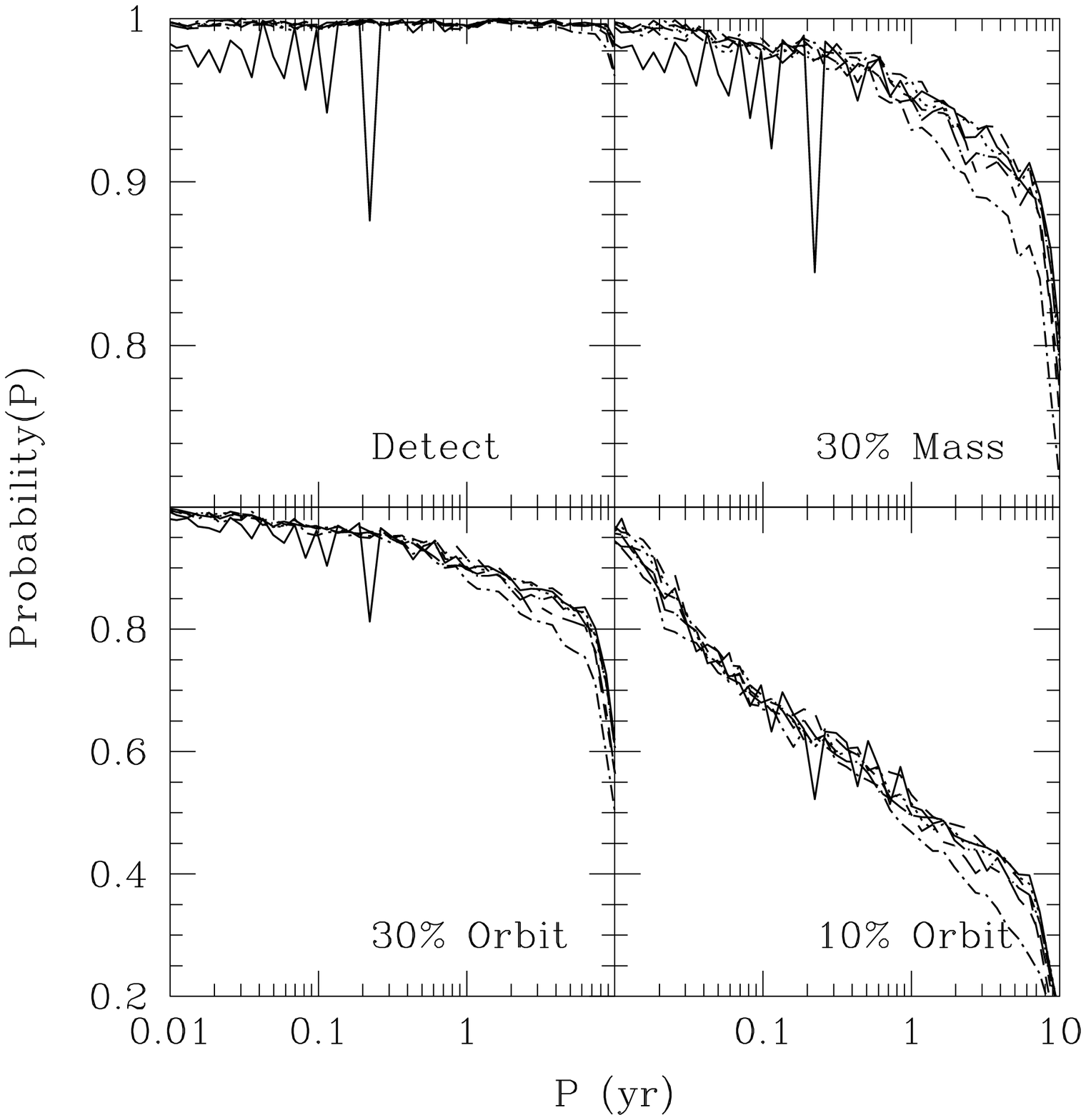}
\caption[fig3.ps]{
In each panel we show the probability for detecting a planet or
measuring its mass or orbit as a function of the orbital period, $P$, for a
fixed scaled signal, $S=6$, averaging over the eccentricities, and
other parameters.  The upper left panel is for detecting a planet, the
upper right panel is for measuring the mass with 30\% accuracy, the
lower left panel is for measuring the orbit with 30\% accuracy, and
the bottom right panel is for measuring the orbital parameters with
10\% accuracy.  The different line styles are for simulations using
different observing schedules (solid, periodic; dotted, periodic with
perturbations ($\epsilon = 0.2$); dot-long dash, logarithmic ($\Delta
t_{min}=30$ d); short dash, power law ($\beta = 0.5$); dot-short dash,
geometric ($\Delta t_{min}=30$ d); short dash-long dash, Golomb ruler).
There are significant
differences in the efficiency for measuring the masses and orbits of
planets with orbital periods approaching the duration of the survey.
There is also significant aliasing near $P=0.2$ yr (and harmonics of this period) when
using the periodic observing schedule (solid line).
\label{ProbVsPFixedSForBest}}
\end{figure}

In Fig.\ \ref{ProbVsPFixedSForBest}, we plot the rates for detecting
planets and measuring their masses and orbits as a function of orbital
period, while fixing the scaled signal, $S=6$.  Since the scaled
signal is held constant for all orbital periods, the rates are higher
than in Fig.\ \ref{ProbVsLogPForBest} for planets with small orbital
periods, but smaller for planet with long orbital periods.  Note that
there is a sharp decline in rates for measuring masses and orbits for
planets with orbital periods approaching the duration of the
astrometric survey.  The regular periodic (solid line)
and periodic with perturbations (dotted line) observing schedules performed
marginally better for planets with long orbital periods and the geometric
strategy performed worst.

In Fig.\ \ref{ProbVsPFixedSForBest}, the regular periodic observing
schedule (solid line) reveals other features.  The
reduction in the rates of detecting planets and measuring their
masses and orbital periods near $\sim0.2$ yr (and to a lesser extent at
harmonics of this period) for the regular periodic observing schedule is due to
aliasing.  We will address this issue further in \S\ref{SecAliasing}.
Also, note that the regular periodic observing schedule is
significantly less efficient at detecting and measuring the mass of
planets with very small orbital periods, due to the lack of pairs of
observations with small spacings.

\subsection{Rate versus Orbital Eccentricity}

\begin{figure}
\plotone{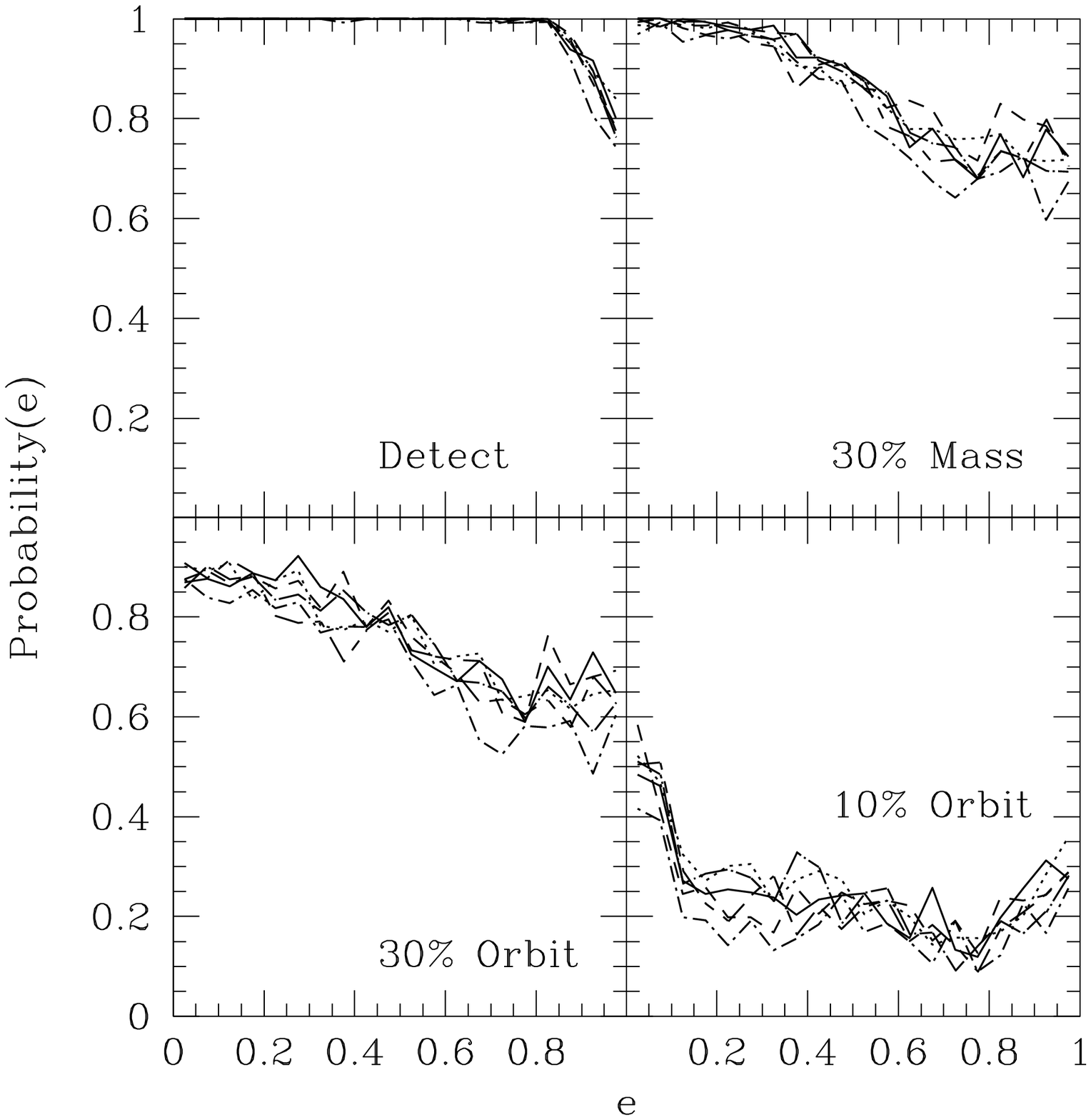}
\caption[fig4.ps]{
In each panel we show the probability for detecting a planet or
measuring its mass or orbit as a function of the orbital eccentricity, $e$, for a
fixed scaled signal, $S=6$, and orbital period, $P=2.5$ yr, averaging over the 
other parameters.  The upper left panel is for detecting a planet, the
upper right panel is for measuring the mass with 30\% accuracy, the
lower left panel is for measuring the orbit with 30\% accuracy, and
the bottom right panel is for measuring the orbital parameters with
10\% accuracy.  The different line styles are for simulations using
different observing schedules (solid, periodic; dotted, periodic with
perturbations ($\epsilon = 0.2$); dot-long dash, logarithmic ($\Delta
t_{min}=30$ d); short dash, power law ($\beta = 0.5$); dot-short dash,
geometric ($\Delta t_{min}=30$ d); short dash-long dash, Golomb ruler).
All of the observing schedules are significantly more efficient for
planets in nearly circular orbits.
\label{ProbVsEFixedSPForBest}}
\end{figure}

In Fig.\ \ref{ProbVsEFixedSPForBest}, we show the rates of detecting
planets and measuring their masses and orbits as a function of orbital
eccentricity for fixed scaled signal, $S=4$, and orbital period,
$P=2.5$ yr.  All the observing strategies which we consider have a
similar functional form and have a significantly lower rate of
characterizing planets with high eccentricities.  As the orbital
period approaches the duration of observations, the effect becomes
significant at smaller eccentricities.  This should be expected due to
projection effects.  For a star perturbed by a single planet, the star
will appear to trace out an ellipse with semi-major and semi-minor
axes proportional to $S \sqrt{1-e^2 \sin^2 \omega}$ and $S \sqrt{1-e^2
\cos^2 \omega} \left|\cos i\right|$.  Planets with large
eccentricities will sometimes be harder to detect (i.e., when the
major axis of the orbit is nearly parallel to the line of sight).

\subsection{Rate versus Inclination}

\begin{figure}
\plotone{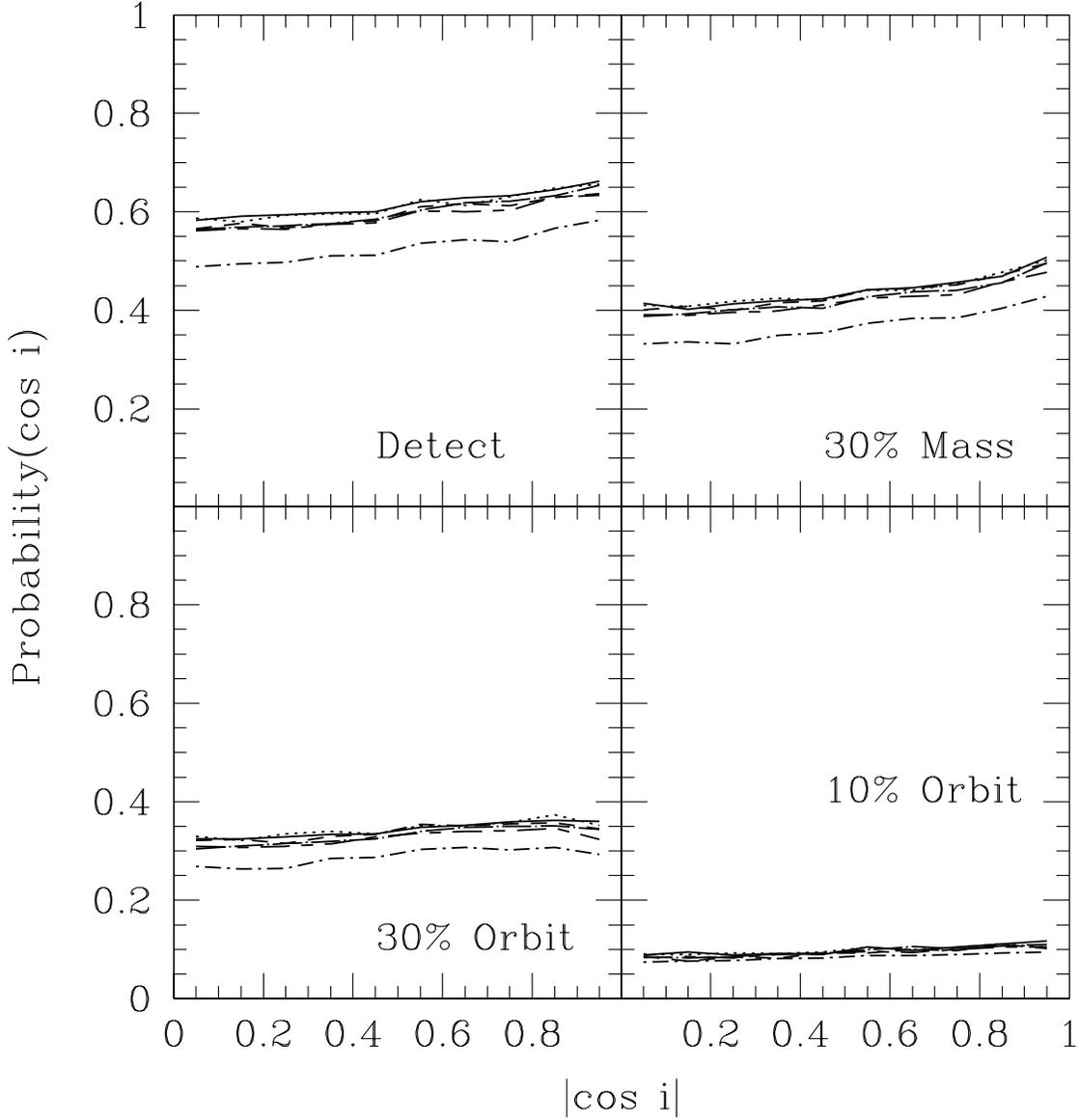}
\caption[fig5.ps]{
In  each panel  we  show the  probability  for detecting  a planet  or
measuring  its mass  or  orbit as  a  function of  the  cosine of  the
inclination to  the plane  of the  sky, $\cos i$,  for a  fixed scaled
signal, $S=2$,  averaging over the  other parameters.  The  upper left
panel  is  for  detecting a  planet,  the  upper  right panel  is  for
measuring the  mass with  30\% accuracy, the  lower left panel  is for
measuring the orbit with 30\%  accuracy, and the bottom right panel is
for  measuring  the  orbital   parameters  with  10\%  accuracy.   The
different line  styles are  for simulations using  different observing
schedules  (solid,  periodic;   dotted,  periodic  with  perturbations
($\epsilon  = 0.2$); dot-long  dash, logarithmic  ($\Delta t_{min}=30$
d); short dash,  power law ($\beta = 0.5$);  dot-short dash, geometric
($\Delta t_{min}=30$ d); short  dash-long dash, Golomb ruler).  All of
the observing  schedules are slightly more efficient  for planets with
small inclinations ($|\cos i| \simeq 1$).
\label{ProbVsCosIFixedSForBest}}
\end{figure}

The efficiency of detecting and characterizing planets also depends on
the inclination of the planet's orbit relative to the plane of the sky
(Sozzetti \etal 2001, 2002, 2003; Eisner \& Kulkarni 2001, 2002).  If
a planet's orbit is exactly edge-on, then the projected motion of the
star is confined to one dimension.  When the baseline of the
interferometer is perpendicular to the orbital plane, observations
provides no information about the orbit.  If the projected orbital
plane and interferometer baseline differ by an angle, $\theta$, then
the amplitude of the signal is reduced by a factor $|\cos i|$.  In
Fig.\ \ref{ProbVsCosIFixedSForBest}, we show the rates for detecting
planets and measuring their masses and orbits as a function of the
inclination for a fixed scaled signal, $S=2$, averaging over the other
parameters.  As expected, all the observing strategies which we
consider have a similar functional form and have a slightly lower rate
of characterizing planets which are viewed nearly edge-on (small $\cos
i$).  The effect becomes less significant for larger scaled signals.
This should be expected due to projection effects.  For small scaled
signal, the inclination effect can result in a significantly reduced
efficiency for detecting a planets for a wide range of orbital
inclinations.  When the scaled signal is large, the effect is
significant only for orbits which are very close to $\cos i = 0$.

\subsection{Regular versus Random Observing Schedules}

\begin{figure}
\plotone{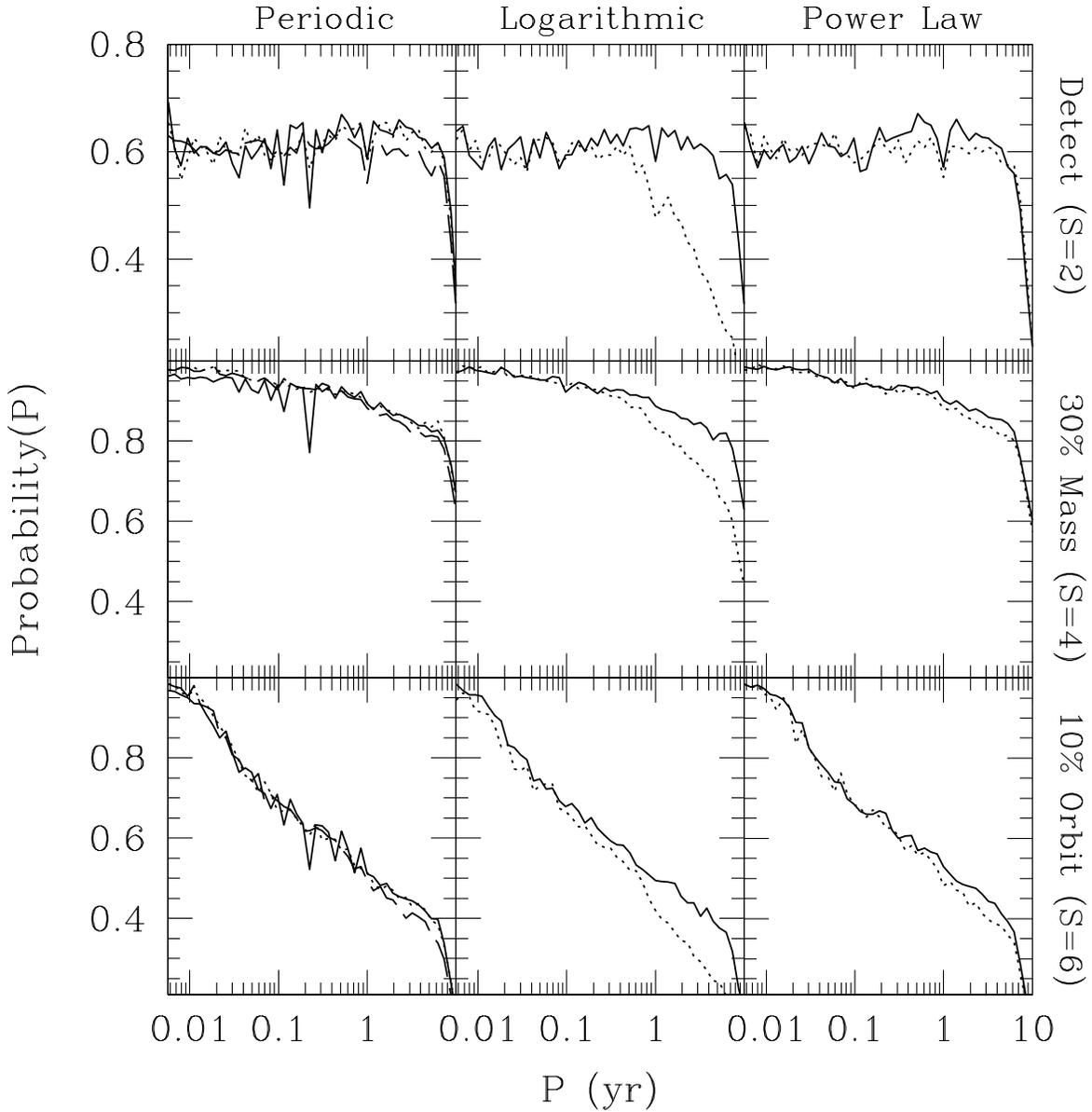}
\caption[fig6.ps]{
In each panel we show the probability for detecting a planet or
measuring its mass or orbit as a function of the orbital period, $P$,
for a fixed scaled signal, $S$, averaging over the other parameters.
The left column is for regular periodic (solid), random uniform
(dashed), and periodic with perturbations ($\epsilon=0.2$, dotted)
observing schedules.  The middle column is for regular logarithmic
(solid) and random logarithmic (dotted) observing schedules with
$\Delta t_{min}=30$ d.  The right column is for regular power law
(solid) and random power law (dotted) with $\beta=0.5$.  The top row
is for detecting the planet (assuming $S=2$), the middle row is for
measuring the planetary mass to with 30\% accuracy (assuming $S=4$),
and the bottom row is for measuring the planet's orbit with 10\%
accuracy (assuming $S=6$).  In nearly all cases, the observing
schedules based on random observing times perform slightly less well
for orbital periods approaching the duration of the astrometric
survey.  The notable exception is for the periodic with perturbations
schedule (left column, dotted line), which performs as well as the
regular periodic schedule for large orbital periods and better for
orbital periods where aliasing would reduce the efficiency of a survey
using the regular periodic schedule.

\label{ProbVsPFixedSRandComp}}
\end{figure}

Next, we compare regular and random observing schedules.
In Fig.\ \ref{ProbVsPFixedSRandComp} we show the rates of detecting
planets and measuring their masses and orbits as a function of orbital
period for fixed $S$.  The solid lines show the results for observing
schedules with regular spacings and the dotted lines show the results
for observing schedules with random observing times drawn from a
distribution based on the regular observing schedule shown in the same
panel.  The left column is for regular periodic (solid), random uniform
(dashed), and periodic with perturbations ($\epsilon=0.2$, dotted)
observing schedules.  The middle column is for regular logarithmic (solid) and
random logarithmic (dotted) observing schedules with $\Delta
t_{min}=30$ d.  The right column is for regular power law (solid) and random
power law (dotted) with $\beta=0.5$.  The top row is for detecting the
planet (assuming $S=2$), the middle row is for measuring the planetary
mass with 30\% accuracy (assuming $S=4$), and the bottom row is
for measuring the planet's orbit with 10\% accuracy (assuming $S=6$).

In most cases, the random version of the observing schedules performs
slightly less well than the similar regular observing schedule,
especially for orbital periods approaching the duration of the
astrometric survey.  This is typical because the random
schedules sometimes contain a larger gap between observations
than would occur in the similar regular observing schedule.

One notable exception occurs for the regular periodic observing
schedule (left column, solid line).  Since regular periodic spacings
result in significant aliasing, adding randomness improves the
efficiency.  However, a uniform random schedule (left column, dashed
line) also suffers somewhat at long orbital periods, as described
above.  An alternative solution to the aliasing problem is to
construct an observing schedule by starting with equal periodic
spacings and then adding a Gaussian perturbation to each observing
time (left column, dotted line).  We explore this possibility further
in \S\ref{SecAliasing}.

\subsection{Rate versus Minimum Spacing}

\begin{figure}
\plotone{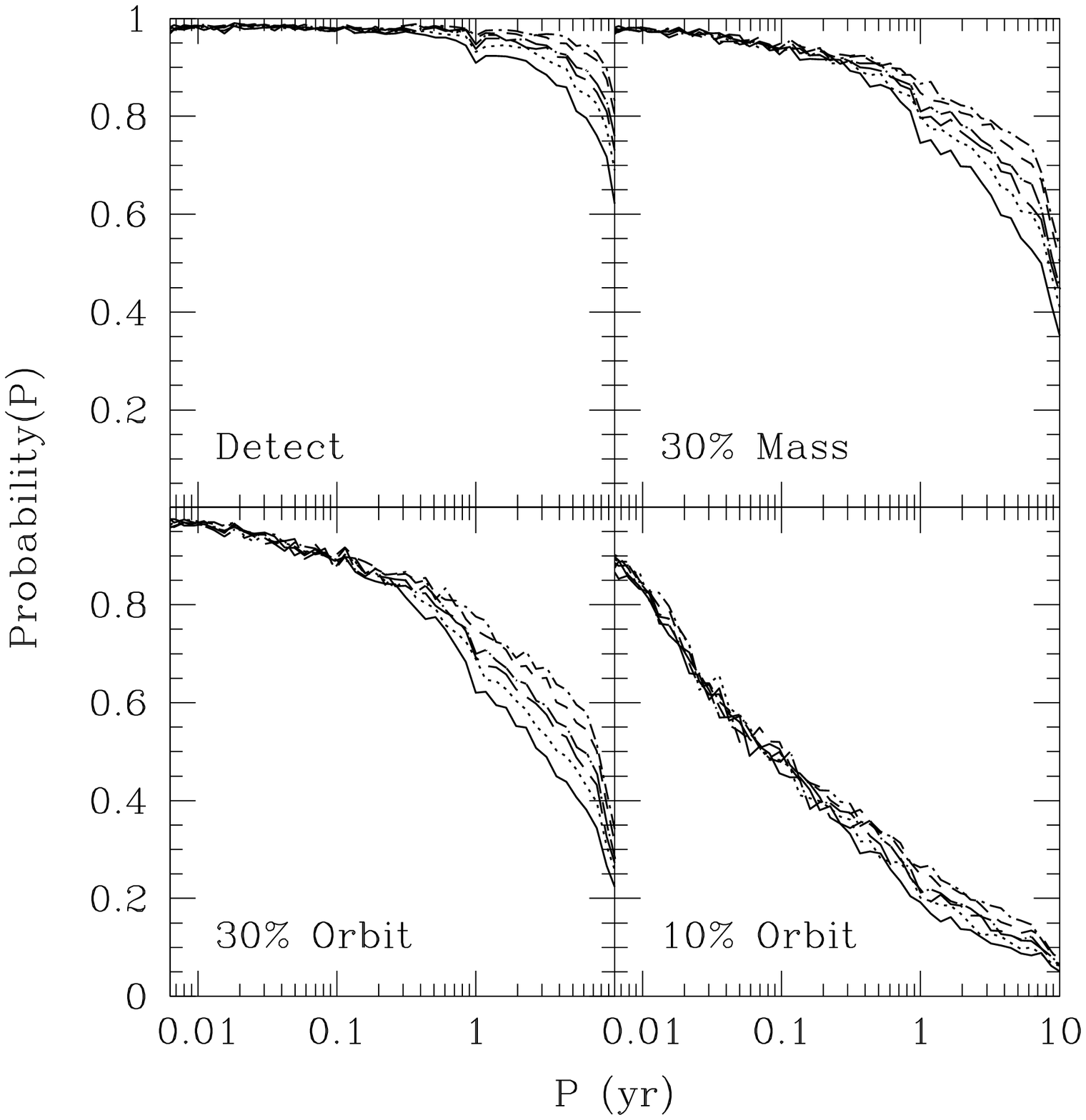}
\caption[fig7.ps]{
In each panel we show the probability for detecting a planet or
measuring its mass or orbit as a function of the orbital period, $P$,
for a fixed scaled signal, $S=4$, averaging over the eccentricities,
and other parameters.  The upper left panel is for detecting a planet,
the upper right panel is for measuring the mass with 30\% accuracy,
the lower left panel is for measuring the orbit with 30\% accuracy,
and the bottom right panel is for measuring the orbital parameters
with 10\% accuracy.  The different line styles are for simulations
using geometric observing schedules with different minimum spacings,
$\Delta t_{min}$ (solid, 1d; dotted, 3d; long dash, 6d; dot-long dash,
10d; short dash, 30d; dot-short dash, 60d).  For orbital periods
approaching the duration of the astrometric survey, the observing
schedules with larger minimum spacings perform better than those will
smaller minimum spacings.  A similar effect occurs for logarithmic and
power law observing schedules.
\label{ProbVsPFixedSForGeo}}
\end{figure}

For each type of regular observing schedule that we considered, the
rates of detecting planets and measuring their masses and orbits
increased as the minimum spacing was increased, and the observing
schedule became more similar to the schedule with regular periodic
spacings.  As an example of this effect, we show the results for
observing schedules using geometric observing schedules with different
minimum spacings in Fig.\ \ref{ProbVsPFixedSForGeo}.  The upper left
panel is for detecting a planet, the upper right panel is for
measuring the mass with 30\% accuracy, the lower left panel is for
measuring the orbit with 30\% accuracy, and the bottom right panel is
for measuring the orbital parameters with 10\% accuracy.  The
different line styles are for simulations using geometric observing
schedules with different minimum spacings, $\Delta t_{min}$ (solid,
1d; dotted, 3d; long dash, 6d; dot-long dash, 10d; short dash, 30d;
dot-short dash, 60d).

We find that there is reduced efficiency for observing schedules with
small $\Delta t_{min}$, consistant with the findings of Sozzetti \etal
(2002).  Since the total number of observations and the survey
duration are held constant, observing schedules with some closely
spaced observations (small $\Delta t_{min}$) also result in some large
gaps between other observations.  This reduces the sensitivity to
planets with orbital periods approaching the duration of the survey.
There is a similar effect for observing schedules using logarithmic or
power law spacings between observations.

\subsection{Aliasing}
\label{SecAliasing}

An observing schedule with constant spacing between observations
(Fig..\ 8, solid line) will obviously have difficulty detecting planets whose
orbital period is nearly equal to the time between observations.  Yet,
observing with a uniform spacing ($\Delta t$) between observations
gives one of the highest overall rates of planet detection.  A variant
on this is an observing schedule in which the observing times are
perturbed by a normal random variable (with standard deviation equal
to $\epsilon \Delta t$).

\begin{figure}
\plotone{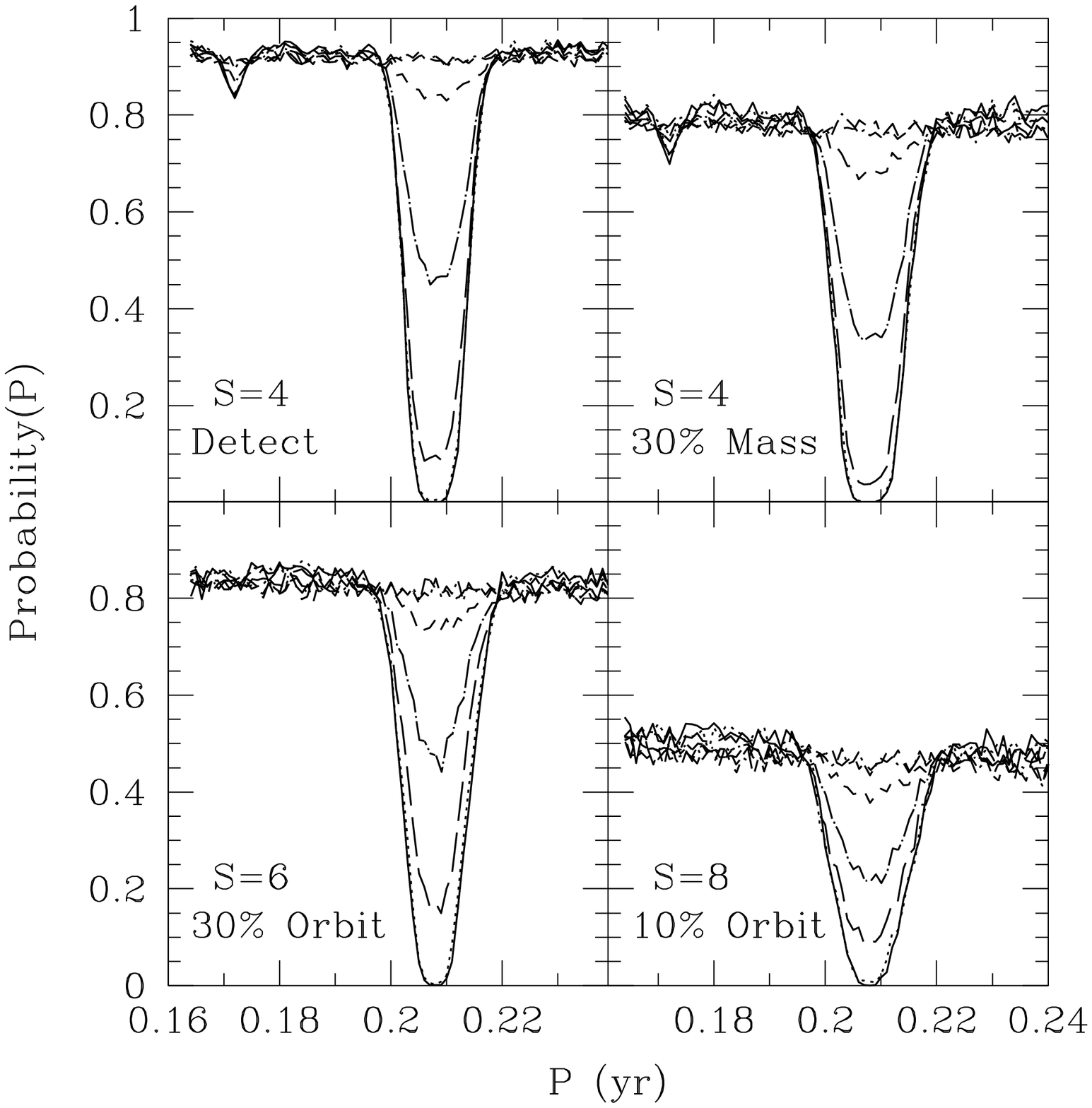}
\caption[fig8.ps]{
In each panel we show the probability for detecting a planet or
measuring its mass or orbit as a function of the orbital period, $P$,
for a fixed scaled signal, $S$, averaging over the other parameters.
The upper left panel is for detecting a planet ($S=4$), the upper
right panel is for measuring the mass with 30\% accuracy ($S=4$), the
lower left panel is for measuring the orbit with 30\% accuracy
($S=6$), and the bottom right panel is for measuring the orbital
parameters with 10\% accuracy ($S=8$).  The different line styles
are for simulations using perturbations with different magnitudes, or
values of $\epsilon$ (solid, $\epsilon=0$, regular periodic; dotted,
$\epsilon=0.02$; dashed, $\epsilon=0.05$; dot-long dash,
$\epsilon=0.1$; short dash, $\epsilon=0.2$; dot-short dash,
$\epsilon=0.4$; short dash-long dash, $\epsilon=0.6$).  Perturbations
with magnitudes $\epsilon\ge0.1$ produce observing schedules which
significantly reduce or even eliminate the aliasing problems
associated with an unperturbed periodic observing schedule (solid
line)  
\label{ProbVsPFixedSForPerGausAlias}}
\end{figure}

In Fig.\ \ref{ProbVsPFixedSForPerGausAlias}, we show the rates of
detecting planets and measuring their masses and orbits as a function
of orbital period for fixed $S$.  Here we have zoomed in to examine
more closely rates for detecting and characterizing planets with
orbital periods near $10 \mathrm{yr}/48 \simeq 0.2$ yr, where aliasing is most significant.
The harmonic at $0.1$ yr can also be seen in this figure.  The solid
lines are for a regular periodic observing schedule, while the other
lines are for a periodic with perturbations observing schedule with
different values of $\epsilon$.  The upper left panel is for detecting
a planet ($S=4$), the upper right panel is for measuring the mass with 30\%
accuracy ($S=4$), the lower left panel is for measuring the orbit with 30\%
accuracy ($S=6$), and the bottom right panel is for measuring the orbital
parameters with 10\% accuracy ($S=8$).

In each panel, the periodic with perturbations observing schedule
significantly reduces the severity of the aliasing for $\epsilon \ge
0.1$ (dot-long dash line) and virtually eliminated the effect for
$\epsilon \simeq 0.4$ (dot-short dash line).  This results in an
slightly higher overall rate of planet detections, and makes the
periodic with perturbations observing schedule the best schedule that
we have examined.

\subsection{Parallax Effect}

Parallax causes nearby stars to appear to trace out an
ellipse on the sky similar to (but typically much larger than) the
perturbation caused by a planetary mass companion.  Since the period
of the parallax effect is one year, planets with
orbital periods nearly equal to one year are more difficult to
detect (Lattanzi \etal 2000; Sozzetti \etal 2002).  

\begin{figure}
\plotone{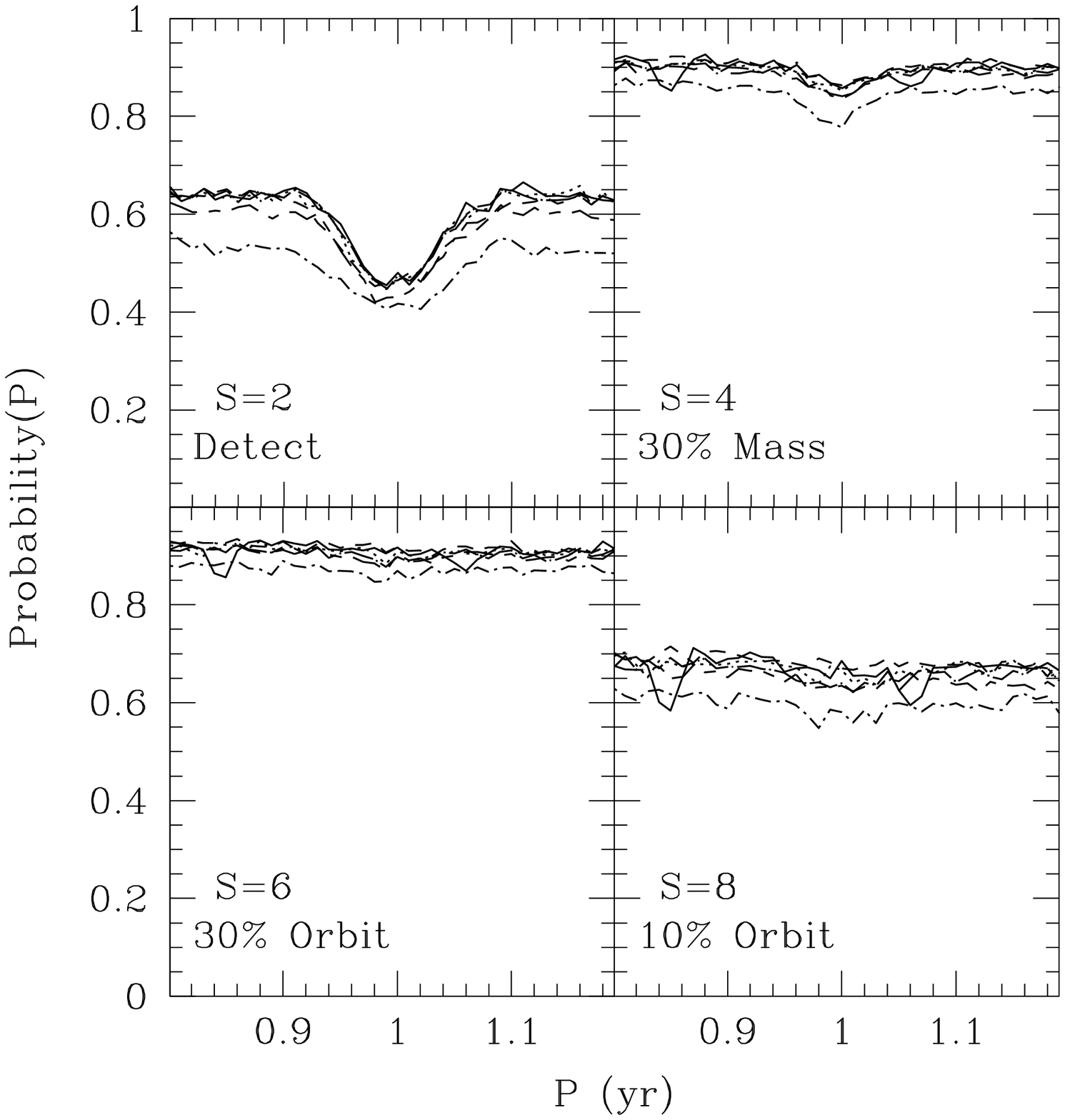}
\caption[fig9.ps]{
In each panel we show the probability for detecting a planet or
measuring its mass or orbit as a function of the orbital period, $P$,
for a fixed scaled signal, $S$, averaging over the other parameters.
The upper left panel is for detecting a planet ($S=2$), the upper
right panel is for measuring the mass with 30\% accuracy ($S=4$), the
lower left panel is for measuring the orbit with 30\% accuracy
($S=6$), and the bottom right panel is for measuring the orbital
parameters with 10\% accuracy ($S=8$).  The different line styles are
for simulations using different observing schedules (solid, periodic;
dotted, periodic with perturbations ($\epsilon = 0.2$); dot-long dash,
logarithmic ($\Delta t_{min}=30$ d); short dash, power law ($\beta =
0.5$); dot-short dash, geometric ($\Delta t_{min}=30$ d); short
dash-long dash, Golomb ruler).  The choice of observing schedule is
not able to eliminate the reduction in efficiency for detecting planets
with orbital periods near $1$ year.  
\label{ProbVsPFixedSForBestParallax}}
\end{figure}

In Fig.\ \ref{ProbVsPFixedSForBestParallax}, we show the
rates of detecting planets and measuring their masses and orbits as a
function of orbital period for fixed $S$.  We have zoomed in to examine
more closely planets with orbital periods near $1$ yr.  
The upper left panel is for detecting a planet ($S=2$), the upper
right panel is for measuring the mass with 30\% accuracy ($S=4$), the
lower left panel is for measuring the orbit with 30\% accuracy
($S=6$), and the bottom right panel is for measuring the orbital
parameters with 10\% accuracy ($S=8$).  The different line styles
are for simulations using different observing schedules.

While there is a small but significant decrease ($\sim30\%$ for $S=2$)
in the detection rate near $P\simeq 1$ year, the rate for measuring
masses or orbits with 10\% accuracy is only slightly affected.  The
choice of observing schedule does not significantly change the
parallax effect.

\section{Discussion}

We have considered several possible observing schedules for a targeted
astrometric planet search similar to SIM.  For most reasonable
observing schedules, the efficiency of planet searches is relatively
insensitive to the observing schedule.  Note that we have used local
fitting algorithms which rely on good initial estimates of the
astrometric and orbital parameters and have not explored possible
problems associated with converging to the correct global solution.

Observing strategies which do not include observations with spacings
less than a planet's orbital period have some difficulty measuring the
orbital parameters accurately.  However, since the amplitude of the
astrometric perturbations scales with $P^{2/3}$, any astrometric planet
search will have trouble detecting very short period planets in any case.

Since the number of observations is held constant, observing schedules
which concentrate too many of their observations in a short period of
time have difficulty accurately measuring orbits for longer period
systems.  This effect favors observing schedules which avoid large
gaps between observations.  All of the regular observing schedules
which we consider can be highly competitive, provided that the minimum
spacing is sufficiently large ($\ge 20$ d).  For these observing
schedules, the minimum spacing is only a factor of a few less than the
average spacing between observations ($\simeq76$ d $\simeq 0.2$ yr).  

Observing schedules with observing times drawn randomly from a broad
probability distribution typically resulted in slightly poorer rates
for detecting and characterizing planets with orbital periods
approaching the duration of the astrometric survey.  This is a result
of occasional large gaps between observing times.  However, drawing
all observation times randomly from a uniform distribution was only
slightly less efficient than the best performing observing schedules
that we considered.

While aliasing is very strong for an observing schedule with regular
periodic observations, the aliasing can be significantly reduced or
virtually eliminated by applying a Gaussian perturbation to the
observing times with fractional standard deviation $\sim10-40\%$ (for
$N_{obs}=48$).  This also improves the rates of detection and
characterization for planets with small orbital periods.  The
magnitude of the perturbation necessary is expected to scale as
$\sim1/N_{obs}$ based on Fourier theory.

Planet searches will also be less sensitive to planets with orbital
periods nearly equal to a year due to the parallax effect.  The fact
that this effect occurs near orbital periods of $\simeq1$ year is
unfortunate, since this is near the range of orbital periods of
particular interest due to the possibility of planets in the habitable
zone of another solar-type stars.  However, for stars of other
spectral types (and hence luminosities), the habitable zone is
expected to occupy different orbital periods, while the parallax
effect always occurs near orbital periods of one 
year.  In any case,
the choice of observing schedule does not significantly reduce the
significance of the parallax effect.

Of the observing schedules which we have considered, we find there is
a small advantage in terms of the rate of detecting planets and
measuring their orbits for an observing schedule obtained by starting
with an equal spacing between observations and adding a small
perturbation ($\simeq 0.4$) to each observing time.  Note that all of
the observing schedules which we have considered are fixed in advance
of all observations.  We have performed additional calculations for $2
N_{obs} = 48$ and verified that our conclusions for the efficiency of
one observing schedule relative to another were not significantly
changed.  In the future, we hope to study the potential benefits of
algorithms which incorporate knowledge gained from previous
observations.  In particular, preliminary results suggest that
significant increases in efficiency are possible by allowing the
number of observations of each target to varry depending on the
outcome of the previous observations.  If it is also
possible to choose the time of additional observations, then even
further improvements in efficiency are likely possible (Ford 2005).

Finally, we emphasize that the differences between observing schedules
which we have identified are relatively small.  This conclusion has
significant implications for planning an astrometric planet search
such as SIM.  Since the exact scheduling of observation times is not
critical, it may be advantageous to schedule observations so as to
minimize time associated with mission overhead (e.g., measuring grid
and reference stars, slewing), allowing a greater number of
observations per target star or a greater number of stars to be
surveyed.

\acknowledgments

We thank Scott Tremaine for his guidance and an anonymous referee for their suggestions.
This research was supported in part by NASA grant NAG5-10456, the
EPIcS SIM Key Project, and the Miller Institute for Basic Research.


\clearpage

\begin{deluxetable}{lcc@{\qquad}cc@{\qquad}cc}
\tablecaption{Rates of Detections and Measurements for All Planets\label{TableSchedStats}}
\tablehead{
   & \multicolumn{2}{c}{Detections} & \multicolumn{2}{c}{Masses} & \multicolumn{2}{c}{Orbits} \\ 
   & $\chi^2$-test & F-test & 30\% & 10\% & 30\% & 10\%
}
\startdata
{\bf Regular Periodic} & 0.66 & 0.59 & 0.62  &  0.54  & 0.58  & 0.46  \\ 
{\bf Random Uniform}  & 0.66 & 0.59 & 0.61  &  0.53  & 0.57  & 0.45  \\ 
{\bf Golomb Ruler}   & 0.66 & 0.59 & 0.61  &  0.53  & 0.57  & 0.45  \\ 
\multicolumn{7}{l}{\bf Periodic w/ Gaussian Perturbations } \\
\qquad $\epsilon = 0.05$  & 0.66 & 0.59 & 0.62  &  0.54  & 0.58  & 0.46  \\ 
\qquad $\epsilon = 0.1$  & 0.66 & 0.59 & 0.62  &  0.54  & 0.58  & 0.46  \\ 
\qquad $\epsilon = 0.2$  & 0.66 & 0.59 & 0.62  &  0.54  & 0.58  & 0.46  \\ 
\qquad $\epsilon = 0.4$  & 0.66 & 0.59 & 0.62  &  0.54  & 0.58  & 0.46  \\ 
%
\multicolumn{7}{l}{\bf Regular Logarithmic } \\
\qquad $\Delta t_{min} = 1$ d   & 0.63 & 0.56 & 0.59  &  0.51  & 0.55  & 0.43  \\ 
\qquad $\Delta t_{min} = 3$ d     & 0.64 & 0.57 & 0.60  &  0.52  & 0.56  & 0.44  \\ 
\qquad $\Delta t_{min} = 10$ d    & 0.65 & 0.59 & 0.61  &  0.53  & 0.57  & 0.45  \\ 
\qquad $\Delta t_{min} = 30$ d    & 0.66 & 0.59 & 0.62  &  0.54  & 0.58  & 0.45  \\ 
%
\multicolumn{7}{l}{\bf Random Logarithmic } \\
\qquad $\Delta t_{min} = 1$ d     & 0.60 & 0.53 & 0.55  &  0.48  & 0.52  & 0.40  \\ 
\qquad $\Delta t_{min} = 3$ d   & 0.61 & 0.54 & 0.56  &  0.48  & 0.52  & 0.41  \\ 
\qquad $\Delta t_{min} = 10$ d  & 0.62 & 0.55 & 0.58  &  0.50  & 0.54  & 0.42  \\ 
\qquad $\Delta t_{min} = 30$ d  & 0.63 & 0.56 & 0.59  &  0.51  & 0.55  & 0.43  \\ 
%
\enddata
\tablecomments{This table lists rates for detecting and characterizing planets with masses $1 M_\oplus$-$10 M_{Jup}$ around
stars at distances of $1-20$pc.  The first column lists the
probability of rejecting the best-fit no-planet model based on a
$\chi^2$-test.  The second column lists the probability that the
best-fit one-planet model significantly reduces $\chi^2$ compared to
the best-fit no-planet model according to an $F$-test.  The remaining
columns list the probabilities of measuring a planet's mass and orbital
parameters to within 30\% and 10\% of their actual values.  The
typical random uncertainty is less than $0.01$.}
\end{deluxetable}

\addtocounter{table}{-1}

\begin{deluxetable}{lcc@{\qquad}cc@{\qquad}cc}
\tablecaption{Cont.}
\tablehead{
   & \multicolumn{2}{c}{Detections} & \multicolumn{2}{c}{Masses} & \multicolumn{2}{c}{Orbits} \\ 
   & $\chi^2$-test & F-test & 30\% & 10\% & 30\% & 10\%
}
\startdata
\multicolumn{7}{l}{\bf Regular Power Law } \\
\qquad $\beta=0.5$   & 0.65 & 0.58 & 0.61 &  0.53 & 0.58 & 0.46 \\ 
\qquad $\beta=1$     & 0.66 & 0.59 & 0.62 &  0.54 & 0.58 & 0.46 \\ 
\qquad $\beta=2$     & 0.66 & 0.59 & 0.62 &  0.53 & 0.58 & 0.45 \\ 
\multicolumn{7}{l}{\bf Random Power Law } \\
\qquad $\beta=0.5$    & 0.65 & 0.58 & 0.61 &  0.53 & 0.57 & 0.45 \\ 
\qquad $\beta=1$      & 0.64 & 0.57 & 0.60 &  0.52 & 0.56 & 0.45 \\ 
\qquad $\beta=2$      & 0.62 & 0.55 & 0.59 &  0.51 & 0.55 & 0.44 \\ 
\multicolumn{7}{l}{\bf Regular Geometric } \\ 
\qquad $\Delta t_{min} = 1$ d    & 0.62 & 0.55 & 0.57 &  0.49  & 0.53  & 0.41  \\ 
\qquad $\Delta t_{min} = 3$ d    & 0.63 & 0.56 & 0.58 &  0.50  & 0.54  & 0.42  \\ 
\qquad $\Delta t_{min} = 10$ d   & 0.63 & 0.57 & 0.59 &  0.51  & 0.55  & 0.43  \\ 
\qquad $\Delta t_{min} = 30$ d   & 0.64 & 0.58 & 0.60 &  0.52  & 0.56  & 0.44  \\ 
%
\enddata
\end{deluxetable}

\begin{deluxetable}{lcc@{\qquad}cc@{\qquad}cc}
\tablecaption{Rates of Detections and Measurements for Low Mass Planets\label{TableSchedStatsTerr}}
\tablehead{
   & \multicolumn{2}{c}{Detections} & \multicolumn{2}{c}{Masses} & \multicolumn{2}{c}{Orbits} \\ 
   & $\chi^2$-test & F-test & 30\% & 10\% & 30\% & 10\%
}
\startdata
{\bf Regular Periodic}  &  0.30 &  0.21 & 0.24 &  0.16 & 0.21 & 0.09 \\ 
{\bf Random Uniform} &  0.29 &  0.20 & 0.23 &  0.15 & 0.19 & 0.08 \\ 
{\bf Golomb Ruler}   &  0.29 &  0.20 & 0.23 &  0.15 & 0.19 & 0.08 \\ 
\multicolumn{7}{l}{\bf Periodic w/ Gaussian Perturbations } \\
\qquad $\epsilon = 0.05$  &  0.30 &  0.21 & 0.24 &  0.16 & 0.20 & 0.09 \\ 
\qquad $\epsilon = 0.1$ &  0.30 &  0.21 & 0.24 &  0.16 & 0.20 & 0.09 \\ 
\qquad $\epsilon = 0.2$  &  0.30 &  0.20 & 0.24 &  0.16 & 0.20 & 0.09 \\ 
\qquad $\epsilon = 0.4$ &  0.30 &  0.20 & 0.24 &  0.16 & 0.20 & 0.09 \\ 
%
\multicolumn{7}{l}{\bf Regular Logarithmic } \\
\qquad $\Delta t_{min} = 1$ d   &  0.25 &  0.16 & 0.18 &  0.11 & 0.15 & 0.06 \\ 
\qquad $\Delta t_{min} = 3$ d  &  0.26 &  0.17 & 0.20 &  0.13 & 0.17 & 0.06 \\ 
\qquad $\Delta t_{min} = 10$ d  &  0.28 &  0.19 & 0.22 &  0.14 & 0.19 & 0.08 \\ 
\qquad $\Delta t_{min} = 30$ d  &  0.30 &  0.20 & 0.23 &  0.15 & 0.20 & 0.09 \\ 
%
\multicolumn{7}{l}{\bf Random Logarithmic } \\
\qquad $\Delta t_{min} = 1$ d   &  0.21 &  0.13 & 0.15 &  0.09 & 0.12 & 0.04 \\ 
\qquad $\Delta t_{min} = 3$ d   &  0.23 &  0.14 & 0.16 &  0.10 & 0.13 & 0.04 \\ 
\qquad $\Delta t_{min} = 10$ d  &  0.24 &  0.15 & 0.18 &  0.11 & 0.14 & 0.05 \\ 
\qquad $\Delta t_{min} = 30$ d  &  0.26 &  0.17 & 0.19 &  0.12 & 0.16 & 0.06 \\ 
%
\enddata
\tablecomments{This table lists rates for detecting and characterizing
planets with masses $1 M_\oplus$-$20 M_\oplus$ around stars at
distances of $1-20$pc.  The typical random uncertainty is less than $0.01$.}
\end{deluxetable}

\addtocounter{table}{-1}

\begin{deluxetable}{lcc@{\qquad}cc@{\qquad}cc}
\tablecaption{Cont.}
\tablehead{
   & \multicolumn{2}{c}{Detections} & \multicolumn{2}{c}{Masses} & \multicolumn{2}{c}{Orbits} \\ 
   & $\chi^2$-test & F-test & 30\% & 10\% & 30\% & 10\%
}
\startdata
\multicolumn{7}{l}{\bf Regular Power Law } \\
\qquad $\beta=0.5$   &  0.30 &  0.20 & 0.24 &  0.16 & 0.20 & 0.08 \\ 
\qquad $\beta=1$     &  0.30 &  0.21 & 0.24 &  0.16 & 0.20 & 0.09 \\ 
\qquad $\beta=2$     &  0.29 &  0.20 & 0.23 &  0.15 & 0.20 & 0.08 \\ 
\multicolumn{7}{l}{\bf Random Power Law } \\
\qquad $\beta=0.5$   &  0.29 &  0.20 & 0.23 &  0.15 & 0.19 & 0.08 \\ 
\qquad $\beta=1$     &  0.29 &  0.19 & 0.23 &  0.15 & 0.19 & 0.07 \\ 
\qquad $\beta=2$   &  0.28 &  0.18 & 0.22 &  0.14 & 0.18 & 0.06 \\ 
\multicolumn{7}{l}{\bf Regular Geometric } \\
\qquad $\Delta t_{min} = 1$ d   &  0.23 &  0.14 & 0.17 &  0.10 & 0.13 & 0.05 \\ 
\qquad $\Delta t_{min} = 3$ d   &  0.24 &  0.15 & 0.18 &  0.11 & 0.15 & 0.05 \\ 
\qquad $\Delta t_{min} = 10$ d  &  0.26 &  0.17 & 0.19 &  0.12 & 0.16 & 0.06 \\ 
\qquad $\Delta t_{min} = 30$ d  &  0.27 &  0.18 & 0.20 &  0.13 & 0.17 & 0.07 \\ 
%
\enddata
%
%
\end{deluxetable}


\end{document}